\documentclass[a4paper,english]{article}
\pdfoutput=1
\usepackage{color}
\usepackage{setspace}

\usepackage[numbers]{natbib}

\definecolor{deeppink}{rgb}{1.0, 0.05, 0.7}

\usepackage{natbib}
\usepackage{latexsym}

\usepackage[T1]{fontenc}
\usepackage[utf8]{luainputenc}
\usepackage{geometry}
\usepackage{amsthm}

\usepackage{amsfonts,amsmath,amssymb}
\usepackage{color}
\usepackage{amsthm}
\usepackage{colortbl}
\usepackage{enumerate}
\usepackage{a4wide}
\usepackage{graphicx}
\usepackage{url}
\usepackage{subfigure}
\usepackage{xcolor}
\usepackage{dsfont}
\usepackage{fancybox}

\usepackage{geometry}
\usepackage{graphicx} 
\usepackage{hyperref}

\numberwithin{equation}{section}
\numberwithin{figure}{section}
  \theoremstyle{plain}
  \newtheorem*{thm*}{\protect\theoremname}

\makeatother

\usepackage{babel}
  \addto\captionscanadian{\renewcommand{\theoremname}{Theorem}}
  \addto\captionsenglish{\renewcommand{\theoremname}{Theorem}}
  \providecommand{\theoremname}{Theorem}

\title{Critical Mathematical Economics \\ and Progressive Computer Science}
\author{Johannes Buchner}

\begin{document}
\maketitle

\begin{abstract}

The aim of this article is to present elements and discuss the potential of a research program at the intersection between mathematics and heterodox economics, which we call Criticial Mathematical Economics (CME). We propose to focus on the mathematical and model-theoretic foundations of controversies in economic policy, and aim at providing an entrance to the literature as an invitation to mathematicians that are potentially interested in such a project.  \\

From our point of view, mathematics has been partly misused in mainstream economics to justify `unregulated markets'. We identify two key parts of CME, which leads to a natural structure of this article: The first part focusses on an analysis and critique of mathematical models used in mainstream economics, like e.g. the Dynamic Stochastic General Equilibrium (DSGE) in Macroeconomics and the so-called ``Sonnenschein-Mantel-Debreu''-Theorems.

The aim of the second part is to improve and extend heterodox models using ingredients from modern mathematics and computer science, a method with strong relation to Complexity Economics. We exemplify this idea by describing how methods from Non-Linear Dynamics have been used in Post-Keynesian Macroeconomics', and also discuss (Pseudo-) Goodwin cycles and possible Micro- and Mesofoundations.\\

Finally, we outline in which areas a collaboration between mathematicians and heterodox economists could be most promising, and discuss both existing projects in such a direction as well as areas where new models for policy advice are most needed. In an outlook, we discuss the role of (ecological) data, AI and the need for what we call Progressive Computer Science.




\end{abstract}

\clearpage

\tableofcontents

\clearpage

\section{Introduction}


The Financial Crisis 2008-2009 has been associated with a crisis in economic theory, or even a ``systemic failure of academic economics'' (\cite{dahlem08}). Although there had been critical voices before that time (e.g. \cite{keen01}, calling economics the ``Naked Emperor of the Social Sciences''), the financial crisis was a turning point with even the general public and journalists asking questions about the unability of the economics profession to foresee such a major crisis. Some even noted ``the unfortunate uselessness of most ’state of the art’ academic monetary economics'' (\cite{buiter09}).

From our point of view, mathematics has been partly misused in mainstream economics to justify `unregulated markets' before the crisis. This can be illustrated by a quote from \cite{grasselli16}: ``Prior to the financial crisis, mathematicians helped to develop the Neoclassical paradigm simply because, to them, it was synonymous with economics. Since then, awareness of non-Neoclassical approaches has spread amongst mathematicians.'' 

As one of the responses to the Financial Crisis, the ``Institute for New Economic Thinking'' (INET \cite{inet09}) emerged, and it has called for a major change in ``Economics after the Crisis'' (\cite{turner12}). It explicitly supports interdisciplinary perspectives (taken from \cite{inet16}):\\

``'The 2008 Crisis demonstrated the need for more diverse approaches to math and economics. Using complexity economics, mathematical physics, and modern dynamical systems theory, we can reach beyond the traditional models ubiquitous in business schools and economics departments today.''\\

One example is the workshop ``Mathematics for New Economics Thinking'' (\cite{grasselli13}), which was an attempt to gather both mathematicians and heterodox economists, organized by Matheus Grasselli at the Fields Institute for Research in Mathematical Sciences at the University of Toronto.  

Also, there has been a ``Workshop on Complexity and Economics'' (\cite{comp-workshop}) at the Fields Institute, which was organized in collaboration with the ``Young Scholars Initiative'' (YSI) of INET (\cite{ysi}) and especially the ``YSI Complexity Economics Working Group'' (\cite{ysi-complexity}. The aim of this article is to describe some of the research is this direction - in order to see what mathematics and mathematicians can contribute to the development of a competitive alternative to the neoclassical paradigm in economics.

\section{The Use of Mathematics in Standard Economics}

\subsection{DSGE models in macroeconomics}

Mainstream macroeconomics is relying heavily on Dynamic Stochastic General
Equilibrium (DSGE)-models, partly because of their microfoundation based
on neo-classical General Equilibrium Theory. Also, these are the type of models that are mostly used in central banks and for policy advice (see e.g. \cite{smets03}). That's why we discuss them here - for a recent survey of their major drawbacks, see e.g. \cite{abm-sfc}, where it is stated that ``DSGE models still fail to recognize the complex adaptive nature of economic systems, and the implications of money endogeneity. `` (\cite{abm-sfc}, p.1)

In the paper \cite{Grasselli-Agent-based},
Grasselli and Ismail articulate two main points of critique of DSGE models and their microfoundation - they state that ``both
‘rational expectations’ are an inherently inappropriate way to make
forecasts under frequent unanticipated changes in the environment
and that models based on ‘representative agents’ merely assume away
the solution of the aggregation problem, entirely disregarding the
powerful negative results on stability and uniqueness of equilibrium
provided by the Sonnenschein-Mantel-Debreu theorems'' (\cite{Grasselli-Agent-based}
p. 2, also see \cite{Sonnenschein72,Mantel,Debreu} and section \ref{sub:Theorem-of-Sonnenschein-Mantel-D} of this paper). 

They propose to use an ``agent-based computational model instead,
which does not rely on any free-floating notion of equilibrium either,
with the possible outcomes being the result of the interactive dynamics
for agents, as is the case with most complex adaptive systems''.
In the paper, they develop such a model in the context of banking
systems. They are confident that ``agent-based
computational models constitute an important new weapon in the arsenal
of statistical, mathematical and economic methods deployed to understand
and mitigate systemic risk in modern banking systems'' (\cite{Grasselli-Agent-based},
p. 25). 

To our knowledge, there is no explicitly ``mathematical critique of DSGE models'', maybe with the exception of \cite{racritique}, which ``draws attention to the problems inherent in the technique of local linearisation and concludes by proposing the use of nonlinear models, analysed globally.''. 

  But there is a considerable literature with critical reflections on DSGE models (see e.g. \cite{colander06}), which is growing after the financial crisis (for further references see \cite{cosma15}). While some mathematicians have focussed on criticizing DSGE-models from a statistical/forecasting point of view (\cite{cosma15}), we choose to focus on a theoretical problem here and elaborate on the aggregation problem in the next section.


\subsection{Theorem of Sonnenschein-Mantel-Debreu on Stability and Uniqueness
of Market Equilibria\label{sub:Theorem-of-Sonnenschein-Mantel-D}}

The theorem of Sonnenschein-Mantel-Debreu has tremendous consequences
and could be seen as the ``endpoint'' of the neoclassical attempt
to show that a market economy is stable and necessarily maximizes
social welfare. Also, it undermines a key feature of the neoclassical
approach to economics: that ``everything happens in equilibrium''.
For this idea to work out, it would be essential to prove existence,
uniqueness and dynamic stability of a market equilibrium - but this fails already in the simplest interesting
example, a pure exchange economy. While the existence of a market
equilibrium can be proved without further assumptions (by using fixed-point
theorems), the theorem of Sonnenschein-Mantel-Debreu is a ``negative
result'' in the way that it shows the complete arbitrariness of the
excess demand functions under the standard assumptions in microeconomics. 

The structure of this section is as follows: We first refer to the original papers but then state and explain a modern version of the theorem. In the final part, we give an outlook how methods from symplectic topology can be used to adress the aggregation problem.

\subsubsection{Original Papers and Recent References}

The now so-called ``Sonnenschein-Mantel-Debreu-theorem'' or ``Sonnenschein-Mantel-Debreu-conditions''
have their origin in a few papers from the seventies (\cite{Sonnenschein72,Sonnenschein73,Mantel,Debreu}).
They use methods from differential topology and global analysis and are demanding to read from a technical point of view. A more accessible treatment can be found in \cite{kirman92}, where the question ``whom or what does the representative individual represent'' is asked. For a recent discussion and further references, see \cite{Rizvi06}.

\subsubsection{A Simplified Modern Version of the Theorem}

The aim of the next section is to present a simplified modern version of the theorem, and explain its mathematical formulation. The theorem and description below is taken (with some simplifications)
from chapter 5 on exchange economies from \cite{Mas-Collel-Book}:

\begin{thm*}
Let $f:S\rightarrow\mathbb{R}^{l}$ be a $C^{3}$- vector field satisfying
some boundary conditions. Then for any $\epsilon>0$ we can find an
economy $E$ such that the excess demand function of $E$ coincides
with $f$ on $S_{\epsilon}$ and p is an equilibrium for $E$ if and
only if $f(p)=0$, i.e. no new equilibrium is added.
\end{thm*}

In his book (\cite{Mas-Collel-Book}), Andreu Mas-Colell proves a
version of the Sonnenschein-Mantel-Debreu-theorem using topological
index theory. For a regular economy, it turns out that to every equilibrium
one can associate in a natural manner an index equal to plus one or
minus one, in such a way that the sum of these indicies is one. In
particular, this implies that an equilibrium exists because the number
of equilibria must be odd. The key insight is, however, that it is
not possible to derive any stronger theorem from the general hypotheses
describing an exchange economy. To get restrictions on the equilibrium
set beyond the ones yielded by the index theorem (like dynamic stability,
for example), it is necessary to consider special classes of economies.

The central result is the theorem above: It tells us that, except
at the boundary, we are dealing with an arbitrary vector field $f(p)$,
and in general, even for a unique equilibrium, it is possible for
the dynamics not to approach this equilibrium from any initial point,
e.g. because of the existence of a stable periodic orbit surrounding
the (unstable) equilibrium.

\subsubsection{Exterior Differential Calculus and Sonnenschein's problem}

In the paper \cite{PaperEDC}, exterior differential calculus is used to address the problem
of characterizing aggregate demand of a market economy. The authors explain that, from a mathematical standpoint, the
ideas of maximization and aggregation have a natural translation in
terms of combination of gradients. 

Specifically, this means that for
a function $X:\mathbb{R}_{+}^{n}\rightarrow\mathbb{R}^{n}$ representing
aggregate behavior, the question is if it can be decomposed as a linear
combination of gradients $D_{p}V^{k}(p)$, where $V^{k},k=1...K$
are functions defined on $\mathbb{R}_{+}^{n}$:

\[
X(p)=\sum_{k=1}^{K}\lambda_{k}(p)D_{p}V^{k}(p)
\]

A natural question, initially raised in \cite{Sonnenschein72} is the
following: what does the above relation imply upon the form of the
function X? In particular, are there testable necessary restrictions
on the aggregate function $X(p)$ that reflect its decomposability
into individual maximizing behaviour? And is it possible to find sufficient
conditions on $X(p)$ that guarantee the existence of a decomposition
of the above type? In the paper, after giving a general introduction to exterior differential
calculus, two powerful theorems due to Darboux and to Cartan/Kähler
are presented, which are in the end used to address Sonnenschein's
problem. More material in this direction can be found in the book \cite{ema},  called ``The economics and mathematics of aggregation''.



\section{The Dynamical Systems Approach to Macroeconomics\label{dynsys}}

The title of this section is taken from a thesis written under the supervision of M. Grasselli (\cite{lima13}). It uses modern mathematical techniques from Dynamical Systems to analyse differential equations arising in macroeconomic models stemming from a Post-Keynesian tradition (see e.g. \cite{gl07}). For an introduction to "Post-Keynesian Foundations" see \cite{lavoie14}, a recent treatment of "Macroeconomics after Kalecki and Keynes" can be found in \cite{hein23}.

\subsection {Post-Keynesian Macro-Models}

Stock-Flow-Consistent Macro Models, based on Post-Keynesian Economic ideas, are described in detail in \cite{gl07}.
In this line of reseach, supported by an INET grant (\cite{grasselli16}), several extensions of the original Goodwin model of the buisness cycle (\cite{Goodwin1967}) have been analysed. The starting point was the one proposed by S. Keen (\cite{keen95}) to model Minsky’s Financial Instability Hypothesis, and was analysed mathematically in \cite{grasselli12}. An extension containing a government sector was treated in \cite{grasselli14}, while inflation and speculation was incoporated into the model and analysed in \cite{grasselli15}.


\subsection{Goodwin and Pseudo-Goodwin Cycles}

\begin{figure} [htbp]
\begin{center}
\includegraphics[height=6cm]{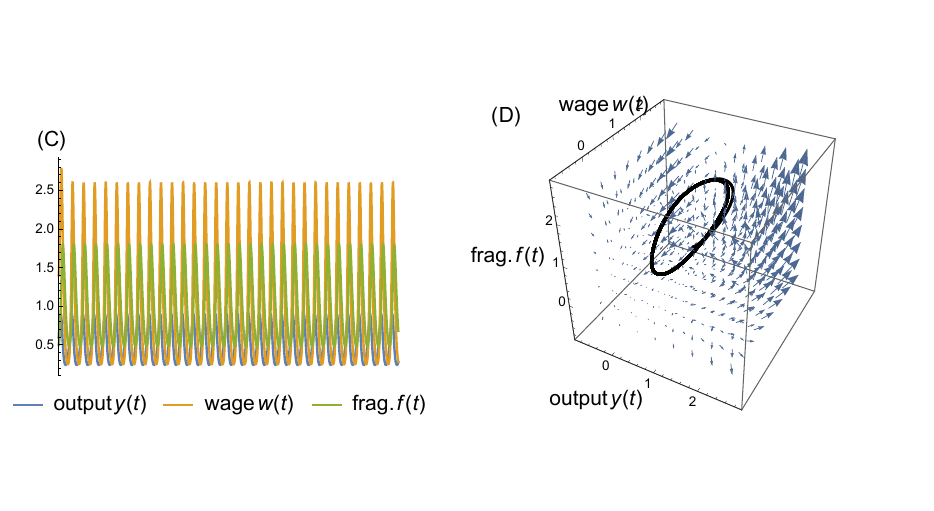}
\caption{Orbit for $s=0$. 
The cycle in $y$ and $f$ drives cyclical behavior in the enslaved variable $w$ ($f$ stands for financial fragility). A pseudo-goodwin-cycle can be observed. This case corresponds to the Minsky model with a reserve army effect. See \cite{brummit16} for a more detailed description and explaination.
}
\label{fig:orbit_s=0}
\end{center}
\end{figure}

The paper \cite{Stockhammer2014} contains material in a similar spirit, studying the interaction of several types of cyclical behaviour in (simplified) Goodwin- and Minsky-models. There, also the relation between Goodwin-cycles and so-called ``Pseudo-Goodwin-cycles'' is discussed, but there is still room to make these concepts more mathematically precise. See \cite{brummit16} for a first step in this direction, including a more detailed analysis of the differential equations and studying the arising Hopf bifurcation. As an illustration, we reproduce one of the figures showing a pseudo-goodwin-cycle (see figure \ref{fig:orbit_s=0}).

All of the works refered to until now in this section deal with the aggregated Macro level of the economy.  The next part of this section will be devoted to the question of a ``Micro- and Mesofoundation'' of Macroeconomics. In \cite{CG}, a ``stochastic microfoundation framework'' for Minsky’s Financial Instability Hypothesis is presented. This approach will also shortly be described in the next section. 



\subsection {A probabilistic Micro-/Meso-foundation of Macroeconomics \label{meso}}

In section \ref{sub:Theorem-of-Sonnenschein-Mantel-D} above, we elaborated on the (shaky) Mircofoundations of DSGE Macro-Models. This section deals with the possibilities of a probabilistic Micro-foundation of Macroeconomics.

In \cite{aoki06}, a perspective from statistical physics and combinatorial stochastic processes towards Marcoeconomics is presented. Macroeconomic models are treated as beeing composed of large numbers of micro-units or agents of several types, and stochastic dynamic and combinatorial aspects of interactions among them are discussed explicitly. For the microeconomic foundations of Macroeconomics by interacting agents, it is stated that `The point is that precise behavior of each agent is irrelevant. Rather we need to recognize that microeconomic behavior is fundamentally stochastic.' (\cite{aoki06}).

In \cite{CG}, more specifically, the aim is construct such a stochastic microfoundation for Minsky’s Financial Instability Hypothesis. The idea ist to identify homogeneous sub-populations of agents, and then to compute mean-field variables and study their interaction. Mathematically, this leads to a master equation: It allows the study of the dynamics of the number of agents in each cluster, i.e. the transition probabilities between the sub-populations.

Such a mean-field approximation represents economic interaction at an intermediate `Meso-level', and is a promising addition to the toolbox. Ultimately, the aim is to combine ingredients from Agent Based Modelling, Statistical Physics and Non-Linear Dynamics in oder to create a unified model of the Micro-, Meso- and Macro-Level of the economy. A first step in this direction can be found in \cite{GGLS}, where a dynamic aggregation of heterogeneous interacting agents leads to what can be called an ``analytical solution for agent based models''. But coupling this to truly non-linear differential equations at the Macro-Level (e.g. of Goodwin-/Keen-type) remains to be done.

  
\newpage

\subsection{Collaboration between Mathematicians and Heterodox Economists}

We have discussed how mathematics is used in the standard model of Macroeconomics and presented some examples of non-mainstream approaches that use tools from advanced mathematics. For a wider discussion how heterodox economics can profit from interations with the mathematical community see \cite{keen09}, a project S. Keen called ``Mathematics for pluralist economics''.

As there has been an enormous progress and breadth of mathematical research in the past decades, those interactions are still at the beginning. Possible interesting topics to look at for a co-operation between heterodox economists and mathematicians include:

\begin{itemize}

\item  Systemic Risk in Financial Networks, Contagion and Financial Crises (\cite{BSW,HCCMS, CD2})

\item Mean Field Games and applications in economics (\cite{LasryLions,CarmonaDelarue, GueantLasryLions})

\item Non-Ergodicity, Entropy and and Ecological Economics (\cite{rosser99,
rosser01, georgescu71})

\item Path Dependency, Hysteresis and Macrodynamics (\cite{hysteresis})

\end{itemize}

A particularly promising example can be found in \cite{lipton15}, where (stochastically modified) Keen-type macroeconomic differential equation are integrated with a model of an interlinked banking system in a dynamic way. This combines ideas from the modelling of systemic risk in banking systems (where a recent reference is \cite{hurd-book}) with the methods described in section \ref{dynsys}.

From a Dynamical Systems point of view, other promising topics to look at include Hysteresis, e.g. with relation to Keynesian modelling of the labour market (\cite{hysteresis2}). In the future maybe even advanced topics like Spatio-Temporal Patterns, Turing instability or Delayed Feedback Control could be part of the mathematics used for new economic thinking - for a panorama of advanced methods in dynamical systems theory and their interplay with a wide range of applications see e.g.  (\cite{patterns}).



\section{Conclusion: Foundations of Controversies in Economic Policy}

Recently, there has been a growing interest in the relation between Complexity Economics and Policy (see e.g. \cite{colander14}), and Wolfram Elsner described the ``Policy Implications of Economic Complexity and Complexity Economics'' (\cite{elsner15}). In this direction, we propose to clarify the mathematical and model-theoretic foundations of controversies in economic policy. The idea is to check the political demands in debates on economic policy and in how far they (explicitly or implicitly) relate to mathematical models. This can illuminate the controversies in and the debates about economic policy by showing in detail which modelling assumption leads to which policy outcome. 

One important pillar of such a project is a didactic presentation of the results, both for the general public as well as for potential ``multipliers'' like students, journalists and policy makers. A first step in this direction is tackled in the Project ``Model-theoretic Foundations of Controversies in Economic Policy - a Computer-based Didactic Presentation'' (\cite{mgwk}) at the Institute for Political Economy of the Berlin School of Economics and Law (\cite{hwr}). The aim of the project is ``to create a didactic tool with elements of a computer game which makes those controversies explicit by presenting competing paradigms in economics and their economic policy conclusions in an interactive way. This should contribute to pluralism in the education and teaching of economics, as well as to a democratization of economic knowledge in the society as a whole.'' (see \cite{mgwk}). 

We also suggest to develop new models for policy advice in areas which are discussed in the general public and in politics, but currently are non-existent or extremely underrepresented in academic economics. Examples include debates about Post-/De-Growth, basic income or gender relations. The latter would require a project we propose to call ``Gender-Aware Economics'', and is discussed e.g. at the Young Scholars Initiative of INET (\cite{gender}), amoung other places. Although there is a community of ``feminist economics'' (see e.g. \cite{fem1}), it is virtually absent in most mainstream economics departments or let alone macroeconomic models used for policy advice. There has also not yet been an attempt to model the economic impact of traditional and progressive male role models (see e.g. in \cite{maenner}).

So the goal of the project we describe is to go beyond the traditional scope of economics, and we suggest to build a model for what has been called ``Solidaric Modern Times'' (in reference to the German think tank ``Institut Solidarische Moderne''  \cite{ism}). This could include trying to model the concept of ``resonance'' from psychology (see \cite{rosa}, p. 71, where the question is asked why we accept to live in collective structures that make it very difficult for many if not most people to have a meaningful and good life). Ultimately, the aim would be to reconcile political reform strategies and more fundamental approaches in a transformative agenda, in order to tackle three important and related questions: Given the power structures that currently exist in the society and the economy, what could and should a progressive government do policywise? Or is it even not advisable to participate in a government and preferable to stay in opposition for left parties in order to shift the balance of power? And, given a majority in the society, how should we organize our collective and economic structures in order to allow ``resonance'' for the highest number if not all people? 


As an example of concrete political demands that we propose to include in such a model is to tackle the question how an "income corridor" with minimum and maximum income could be implemented (e.g. by modelling near 100\% taxation or by putting limits to the amount of money companies can deduct from their tax bills as executive pay). Also, existing (mathematical) models for Macro- and Meso-co-ordination in the economy should be reviewed, and potentially included, in order to build a model of an alternative way to organize the economy, in relation to what is discussed under the headings ``Democratic Socialism'' or ``Economic Democracy" (see e.g. \cite{rosalux}).




Questions like this are discussed intensively in the political left (see e.g. \cite{emali, pragerf, diem25}), but it is not even on the agenda of academic economics departments to think about economic models to answer them (with only a few exceptions, among them e.g. \cite{hwr, bremen, fgw}). In this sense, we agree that there is a ``systematic failure of academic economics'' (\cite{dahlem08}), and we suggest that heterodox economists and critical mathematicians join forces in order to change this situation.




\section{Outlook: Progressive Computer Science}

We are deeply convinced that the ecological crisis necessarily demands research on post-capitalist economic systems, to which the 'STEM subjects'
science, technology, engineering, mathematics (as well as many other fields) urgently need to contribute in order to achieve social progress. So in addition to critical mathematical economics, we propose to do research in what we call 'progressive computer science' (\cite{buchner25a, buchner25b}), with strong connections to (progressive) data science and artificial intelligence. In this last section of this paper, we elaborate a bit more on this promising research area.

The aim for such an ambitious project is e.g. to look at the European economy from an overall, ecological perspective. We propose to evaluate in how far new technologies and AI-methods allow for a new organization of an ecological economy at a European level: Can we move beyond the traditional way of doing
business that all private companies plan only for themselves and achieve a more comprehensive
ecological economic planning at the level of more aggregated entities? Recent developments in Computer Science and AI
(e.g. in the field of reinforcement learning) have enabled economic planning at large scales e.g. in
multinational companies in a way that was unthinkable a decade ago, as it was published in the book "The people's republic of Walmart" \cite{phillips2019walmart}.
We propose to use such state of the art methods from AI, reinforcement learning and game tree search to evaluate the material flows in the European Economy, and
to discuss the potential for ecological economic planning on a European level on that basis. In the EU, a “Circular economy Action Plan” \cite{eu2} was adopted, and one central part of it is to improve the availibility of digital product data. The “EU digital product passport” (see e.g. \cite{eu1} or \cite{eu3}) is designed to
provide information about each product’s origin, materials, environmental impact, and disposal
recommendations. With this European-wide product data infrastructure for all products sold and used in the EU, new forms of ecological
economic planning will be become possible. The discussion about the requirements and design of
digital product passes has just started in recent years, and we propose to define requirements for a digital product
passport to enable European ecological economic planning.

So in the future, a lot of (ecological) data will become available, including ecological data e.g. for so-called “lifecycle assessment”. This creates the need for "Progressive Computer Science", i.e to use methods from computer and data science (and AI) for a progressive change of the economic system. Since a few years, there is a hype around „Artificial Intelligence“ (AI), and we propose to have a closer look on the role that digital product data and AI (in a broad definition, see below) can play for a democratic economic system that respects planetary boundaries. One focus lies naturally on ecological questions, such as those arising in order to „close the loops“ of material flows and achieving a „circular economy“, while also the role of care work and “male role models” is very important and should be considered from the start (see \cite{vonheesen2022} or \cite{lutosch}). The current hype of AI was fuelled by the introduction of large language models (“LLMs”) like ChatGPT, and nowadays “AI” and “LLMs” are often used interchangeably. However, for our purpose, AI can and should be defined much more broadly: A classic definition by Marvin Minsky from the 1960s defined AI as “the science of making machines do things that would require intelligence if done by humans”, and in a recent article on artificial intelligence and modern planned economies \cite{spyros2}, Spyros Samothrakis noted that historic debates and proposals on "Computerised Economic Planning" (CEP, also see \cite{spyros1}) “have been inspired by and/or have touched upon other numerate disciplines like cybernetics, game theory, optimisation, complex systems, machine learning, and statistics. Arguably, the applied cutting edge of the fields that have partially contributed to the debate is currently being studied under the broad umbrella of artificial intelligence (AI)".

Recently, there has been an application of data from "Life cycle sustainability assessment" to planning purposes. A starting point for the project of "Progressive Computer Science" could e.g. be 
openLCA (see openlca.org), a free and open source software for modelling the life cycle of products and sustainability. For more context on LCA, see e.g. \cite{Zeug2023}. Ultimately, we believe a combination of Critical Mathematical Economics and Progressive Computer Science is necessary in order to achieve economic democracy within planetary boundaries, as outlined e.g. in \cite{sorgbuch} or in the final section of \cite{buchner25}.


\newpage

\end{document}